\documentstyle[procl]{article}

\input{psfig.sty}
\def\Journal#1#2#3#4{(#1) {#2} {\bf #3}, #4}

\def\AAp{\em Astron.\ Astrophys.}

\def\ApJ{\em Astrophys.~J.}

\def\ApJSS{\em Astrophys.~J., Suppl.\ Ser.}

\def\ARAaAp{\em Annu.\ Rev.\ Astron.\ Astrophys.}

\def\MNRAS{\em Mon.\ Not.\ R.~Astron.\ Soc.}
\def\Nat{\em Nature\/}

\def\SA{\em Sov.\ Astron.}

\newcommand{\HI}{{\rm H\,\scriptstyle I}}

\newcommand{\RM}{{\rm RM}}
\newcommand\vect[1]{{\mbox{\boldmath $#1$}}}
\newcommand\sfrac[2]{{\textstyle{\frac{#1}{#2}}}}

\newcommand{\kms}{\,{\rm km\,s^{-1}}}

\newcommand{\cm}{\,{\rm cm}}

\newcommand{\gcmcube}{\,{\rm g}\,{\rm cm^{-3}}}

\newcommand{\mkG}{\,\mu{\rm G}}

\newcommand{\kpc}{\,{\rm kpc}}
\newcommand{\p}{\,{\rm pc}}
\newcommand{\yr}{\,{\rm yr}}

\begin{document}

\addtocounter{page}{190}

\markboth{A.\ Shukurov}{Global magnetic structures}

\thispagestyle{plain}

\title{Global Magnetic Structures in Spiral
Galaxies: Evidence for Dynamo Action}

\author{Anvar Shukurov}

\address{Department of Mathematics, University of Newcastle,\\
Newcastle upon Tyne, NE1~7RU, U.K.}

\maketitle

\abstract{Observational evidence for dynamo action in spiral galaxies
is reviewed, and the capabilities of various theories in explaining
the basic features of galactic magnetic fields are discussed.
Mean-field dynamo models appear to be
unique in providing a coherent explanation of a wide variety of
magnetic features in spiral galaxies.\\
\mbox{}\quad
Reliable modelling of global magnetic structures,
such as the magnetic ring in M\,31, requires detailed knowledge of the
rotation curve, the magnitude and radial profile of
turbulent and noncircular systematic
velocities, the scale height of the warm ionized layer,
the total gas density, the turbulent scale and their variations with
galactocentric radius.  More detailed models involving the effects of
the spiral arms on magnetic field require the knowledge of the
arm-interarm contrasts in the above quantities.
}

%--------------------------------------------------------------
\section{Introduction}
Energy density of interstellar magnetic fields is comparable to
kinetic energy density of interstellar turbulence, so magnetic can
fields significantly affect the turbulent motions. Systematic motions
at a speed in excess of 10--$30\kms$ are too strong to be affected by
interstellar magnetic fields. However, regular magnetic fields revealed
by polarized radio emission can be a sensitive tracer of the regular
motions (e.g., Beck et al., 1999).  Theory and observations of galactic
magnetic fields are now advanced enough to provide useful constraints
on the kinematics and spatial structure of interstellar gas.

There are two basic approaches to the origin of global magnetic
structures in spiral galaxies --- one of them asserts that the observed
structures represent a primordial magnetic field twisted by
differential rotation, and the other that it is due to turbulent
dynamo action. The simplicity of the former theory is appealing, but
it fails to explain the strength, geometry and apparent lifetime of
galactic magnetic fields (Ruzmaikin et al., 1988a,b; Beck et al.,
1996; Kulsrud, 1999; see below). Furthermore, there are no mechanisms
known to produce cosmological magnetic fields of required strength
and scale (Beck et al., 1996). Dynamo models appear to be much
better consistent with the observational and theoretical knowledge of
interstellar gas, and almost all models of magnetic fields in
specific galaxies have been formulated in terms of dynamo theory. It
seems to be very plausible that galactic magnetic fields are
generated by some kind of dynamo action, i.e., that they are produced
{\em in situ.\/} The most promising is the mean-field turbulent
dynamo. The aim of this paper is to justify these statements.
MHD density waves (Lou \& Fan, 1996, 1998, 2000) can add
further structure to the global magnetic fields, but this theory
cannot explain the {\em origin\/} of the regular magnetic field and
relies on a background field supported by an unspecified source.

%---------------------------------------------------
\section{Dynamo control parameters}\label{DCP}

Dynamo action is the conversion of kinetic energy of plasma motion
into magnetic energy independently of any external electric currents.
The dynamo needs a weak seed magnetic field to be launched, but it
does not rely on the seed as soon as the amplification has started,
so the process is called self-excitation. The field amplification is
exponential in time.  Without dynamo action, velocity shear can
amplify an external magnetic field only slowly, linearly in time; the
field eventually decays as soon as  reconnection and magnetic
buoyancy overcome the slow amplification.

Two types of turbulent dynamo can be distinguished: one is
the mean-field dynamo briefly described below, and the other is the
so-called fluctuation dynamo. The {\em mean-field dynamo\/}
generates both regular and random magnetic fields. The
{\em fluctuation dynamo\/} produces purely random magnetic fields
with vanishing mean (regular) component and, contrary to a rather
widespread misconception, acts independently of any overall rotation
and $\alpha$-effect.  The only ingredient needed is a random (even
not necessarily turbulent) motion of conducting fluid with a magnetic
Reynolds number exceeding $R_{\rm m,cr}\simeq100$.  The fluctuation
dynamo contributes to turbulent magnetic fields in the interstellar
medium producing magnetic filaments $l\simeq100\,$pc in length and
$lR_{\rm m,cr}^{-1/2}\simeq10\,$pc in thickness (Sokoloff et al.,
1990; Subramanian, 1999).  The distinction between mean-field and
fluctuation dynamos can be viewed as merely a mathematical
convenience, and unified models have been developed (Subramanian,
1999; Brandenburg, 2000).

Mean-field turbulent dynamo action needs two ingredients, rotation
and deviation of the velocity field from mirror symmetry. The latter
arises naturally in a rotating, stratified system such as the
gas disk of a spiral galaxy. Correspondingly, the mean-field dynamo
is controlled by two dimensionless parameters quantifying the
differential rotation and the so-called $\alpha$-effect,
\begin{equation}                \label{RR}
R_\omega=Gh^2/\beta\;, \qquad  R_\alpha=\alpha h/\beta\;,
\end{equation}
where  $r$ is the galactocentric radius,
$\Omega\simeq25\kms\kpc^{-1}$ is the angular velocity of rotation
obtained from the rotation curve, $G=r\,d\Omega/dr$ is a measure of
differential rotation, $h\simeq500\p$ is the scale height of the
ionized layer (presumably, of the warm interstellar medium),
$\beta\simeq\sfrac13lv\simeq0.3\kpc\kms$ is the turbulent magnetic
diffusivity, $\alpha\simeq l^2\Omega/h\simeq1\kms$ is the helical
(mirror-asymmetric) part of the turbulent velocity $v\simeq10\kms$,
and $l\simeq100\p$ is the turbulent scale (see, e.g., Ruzmaikin et
al., 1988a,b). It is possible that magnetic buoyancy plays
significant r\^ole in galactic dynamos; then $\alpha$ is a function
of magnetic field $\vect{B}$ (Parker, 1992; Moss et al., 1999).

In spiral galaxies, the typical values of the control parameters are
$R_\omega\simeq-10$ and $R_\alpha\simeq1$ at $r\ga1$--$2\kpc$, so
$|R_\omega|\gg R_\alpha$ and then the dynamo (the $\alpha\omega$-dynamo)
is controlled by a single parameter known as the {\em dynamo number\/}
\begin{equation}
D=R_\alpha R_\omega\simeq 10\frac{h^2}{v^2}\,
r\Omega\,\frac{d\Omega}{dr} \;.                \label{D}
\end{equation}
Mean-field dynamo action is also possible with rigid
rotation; then it is called $\alpha^2$-dynamo and its control
parameter is $R_\alpha^2$.

This theory describes
self-excitation of a large-scale or regular magnetic field whose
spatial and temporal scales significantly  exceed those of turbulent
motions, $l\simeq100\p$ and $l/v\simeq10^7\yr$, respectively. The
regular field is in fact that magnetic field which produces polarized
radio emission observed at a linear resolution of 1--3\,kpc typical of
the present-day observations of nearby galaxies.

The regeneration ($e$-folding) rate of the regular magnetic field
$\gamma$ is related to the magnetic diffusion time along the smallest
dimension of the gas layer and to the dynamo number (if $|R_\omega|\gg
R_\alpha$). The following expression is acceptable as a rough estimate:
\begin{equation}
\gamma\simeq \frac{\beta}{h^2}
        \left(\sqrt{\frac{|D|}{D_{\rm cr}}}-1\right)
\simeq(1\mbox{--}10)\,\mbox{Gyr}^{-1}
\qquad\mbox{for } |D|\ga D_{\rm cr}\;,        \label{gamma}
\end{equation}
where $D_{\rm cr}$ is a certain {\em critical\/} dynamo number which
weakly depends on the vertical profile of $\alpha$; $D_{\rm
cr}\approx10$ is a reasonable approximation. Magnetic field can be
amplified ($\gamma>0$) if differential rotation and stratification
are strong enough to yield $|D|\geq D_{\rm cr}$. In spiral galaxies,
this condition is normally satisfied out to a large radius; it is
therefore not surprising that regular magnetic fields have been
detected in all galaxies where observations have sufficient
sensitivity and resolution (Wielebinski \& Krause, 1993; Beck et
al., 1996; Beck, 2000).

The steady-state strength of magnetic field generated by the dynamo
is established when the effects of the regular magnetic field on
turbulent motions become significant. This occurs presumably via
suppression (quenching) of the $\alpha$-effect by the Lorentz force.
This corresponds to a balance of the Lorentz force due
to the regular magnetic field and the Coriolis force that causes
deviations from mirror symmetry.  This yields the following estimate
of the steady-state strength of the regular magnetic field produced
by the mean-field dynamo (Ruzmaikin et al., 1988a,b; Shukurov, 1998):
\begin{equation}
B\simeq\left[4\pi\rho v\Omega l
        \left(\frac{|D|}{D_{\rm cr}}-1\right)\right]^{1/2}
\simeq2\mkG
        \left(\frac{|D|}{D_{\rm cr}}-1\right)^{1/2},
%\left(\frac{n}{1\cmcube}\right)^{1/2}\left(\frac{v}{10\kms}\right)^{1/2},
                                        \label{B}
\end{equation}
where $\rho\simeq1.7\times10^{-24}\gcmcube$ is the density of
interstellar gas.

It is now clear what information is needed to construct a useful
dynamo model: all the variables listed after Eq.~(\ref{RR}) and the
gas density. All these parameters are observable, even though their
observational estimates may be incomplete or controversial. Recent
observational progress has allowed to explore the effects of galactic
spiral patterns on magnetic fields (Beck, 2000). The corresponding
dynamo models require knowledge of the arm-interarm contrast in all
the relevant variables (Shukurov \& Sokoloff, 1998).

%-------------------------------------------------
\begin{figure}[tb]\label{M31pitch}
\begin{tabular}{cc}
\psfig{figure=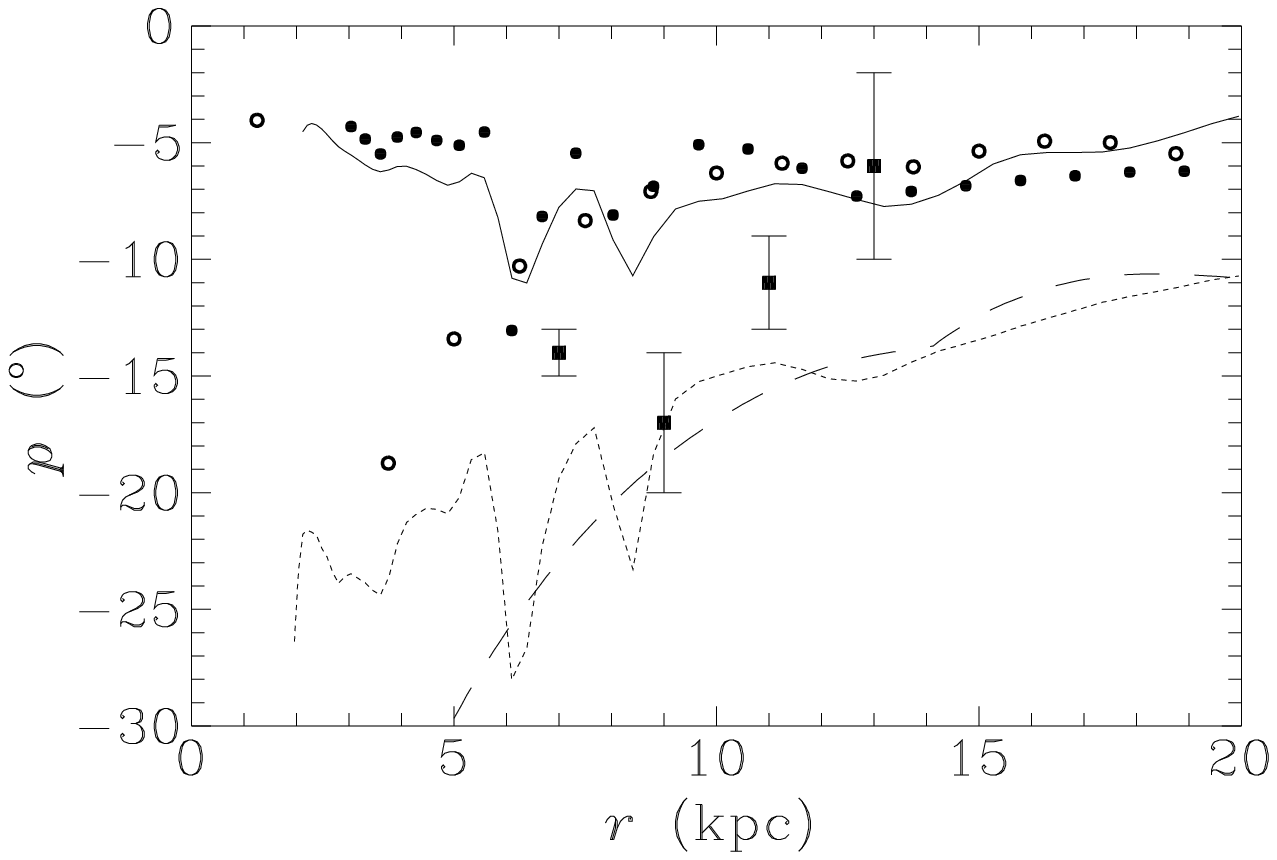,width=7truecm}
&
\psfig{figure=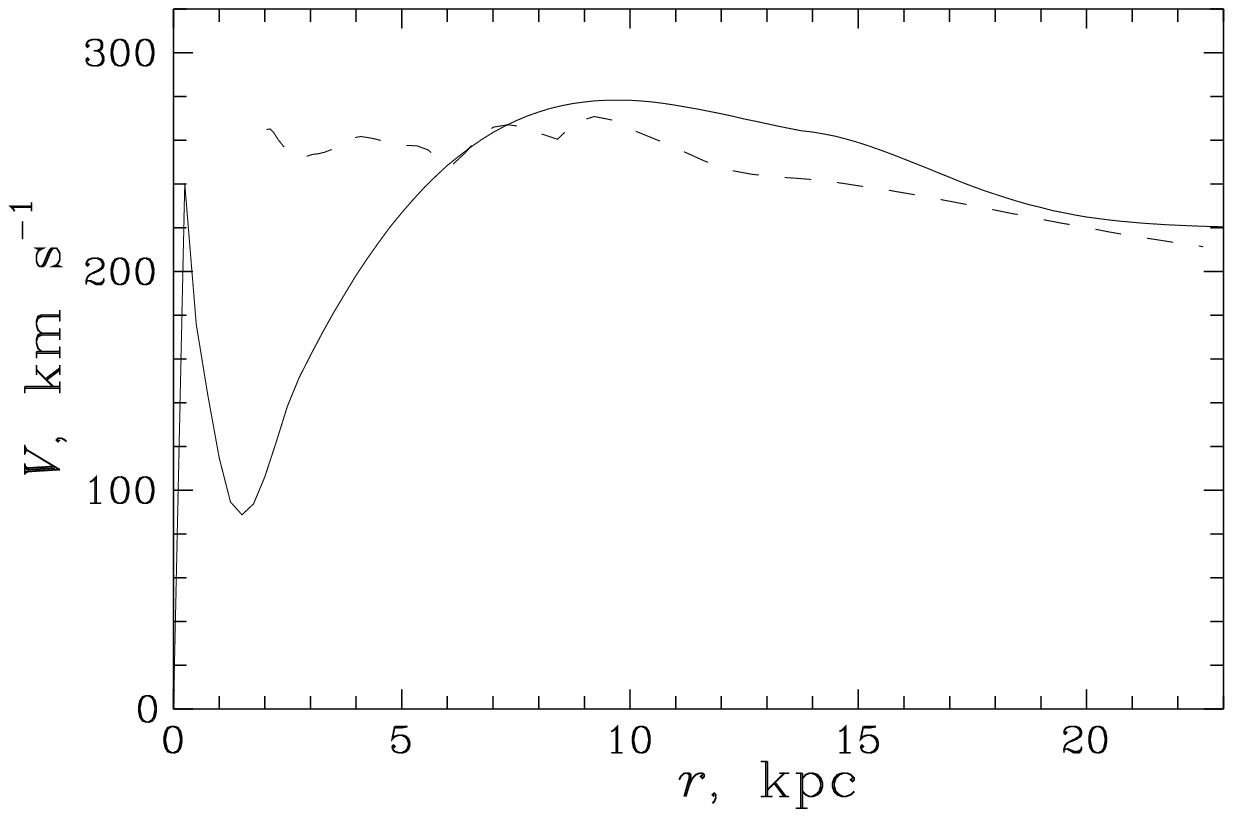,width=7truecm}
\end{tabular}
\caption[]{{\em Left panel:\/} The pitch angle of magnetic field in
M\,31 as obtained from radio polarization observations (squares with
error bars) (Fletcher et al., 2000), from Eq.~(\protect\ref{p}) using the
rotation curve of Deharveng \& Pellet (1975) and Haud (1981)
(dashed) and Braun (1991) (dotted), and from  Eq.~(\protect\ref{pn})
with $D_{\rm cr}=1$ using the same rotation curves (open and filled
circles, respectively); $h(r)$ is twice the $\HI$
scale height of Braun (1991). Results from a nonlinear dynamo model
for M\,31 (Moss et al., 1998) are shown with solid line. {\em Right
panel:\/} The rotation curve of M\,31 from Deharveng \& Pellet (1975)
and Haud (1981) (solid) and from Braun (1991) (dashed). The former is
similar to the rotation curve of Rubin \& Ford (1970).}
\end{figure}
%-----------------------------------------------------------------------------

%---------------------------------------------------
\section{Observational evidence of dynamo action in spiral
galaxies}\label{OEDASG}
%---------------------------------------------------------------------------
\subsection{Magnetic pitch angle}\label{MPA}
Regular magnetic fields observed in spiral galaxies have field lines
in the form of a spiral with a pitch angle in the range
$p=-(10^\circ$--$30^\circ)$, with negative values indicating a
trailing spiral (e.g., Beck et al., 1996).  The value of the pitch
angle is a useful diagnostic of the mechanism maintaining the
magnetic field (Krasheninnikova et al., 1989).

Consider the simplest from of mean-field dynamo equations appropriate
for a thin galactic disk (see Ruzmaikin et al., 1988a,b):
\begin{equation}
\dot B_r=-(\alpha B_\phi)'+\beta B_r''\;,
\qquad
\dot B_\phi=GB_r+\beta B_\phi''\;,
                        \label{dyn}
\end{equation}
where $\vect{B}=(B_r,B_\phi,B_z)$ is the regular magnetic field
written in cylindrical coordinates $(r,\phi,z)$, dot denotes time
derivative, and dash derivative with respect to $z$, the vertical
coordinate; $B_z$ can be obtained, e.g., from
$\nabla\cdot\vect{B}=0$.

Any regular magnetic field maintained by the dynamo must have a non-zero
pitch angle:  for $B_r\equiv0$ (a purely azimuthal magnetic field),
equation for $B_\phi$ in (\ref{dyn}) reduces to a diffusion equation
$\dot B_\phi=\beta B_\phi''$ which only has decaying solutions,
$B_\phi\propto\exp(-\beta t/h^2)$. The same applies to a purely
radial magnetic field.

Consider exponentially growing solutions,
$B_{r,\phi}\propto\exp(\gamma t)$, and replace $\partial/\partial z$
by $1/h$ and $\partial^2/\partial z^2$ by $-1/h^2$ to obtain two
algebraic equations,
$
\left(\gamma+\beta/h^2\right)B_r+\alpha B_\phi/h=0\,,
\
-GB_r+\left(\gamma+\beta/h^2\right)B_\phi=0\,,
$
which have non-trivial solutions only if the determinant vanishes,
which yields $(\gamma+\beta/h^2)^2\simeq-\alpha G/h$, and
Eq.~(\ref{gamma}) follows with $D_{\rm cr}=1$. The resulting
estimate of the magnetic pitch angle is given by
\begin{equation}
\tan p
=\frac{B_r}{B_\phi}
\simeq-\sqrt{\frac{\alpha}{-Gh}}
\simeq-\frac{l}{h}\sqrt{\frac{\Omega/r}{|d\Omega/dr|}}
=-\sqrt{\frac{R_\alpha}{|R_\omega|}}\;.
                                        \label{p}
\end{equation}
For $l/h\simeq1/4$ and a flat rotation
curve, $(\Omega/r)/(d\Omega/dr)=-1$, we obtain $p\simeq-15^\circ$,
and this is the middle of the range observed in spiral galaxies.
More elaborate treatments discussed by Ruzmaikin et al.\ (1988a)
confirm this estimate of $p$ and yield a more accurate value of
$D_{\rm cr}$.

If the steady state is established by reducing $R_\alpha$ to its
critical value as to obtain $R_\alpha R_\omega=D_{\rm cr}$, then
the pitch angle in the nonlinear steady state becomes
\begin{equation}
\tan p\simeq-\frac{\sqrt{D_{\rm cr}}}{|R_\omega|}\;.
                                \label{pn}
\end{equation}

The magnetic pitch angle in M\,31 determined from observations and dynamo
theory is shown in Fig.~1.  Although the model curves show noticeable
differences from the observed pitch angles, the general agreement is
encouraging. The situation is typical: magnetic pitch angles of
spiral galaxies are in a good agreement with predictions of dynamo
theory (Beck et al., 1996).

This picture does not explain why the pitch angles of galactic
magnetic fields are invariably close (though not equal) to those of
the spiral pattern in the parent galaxy. A plausible explanation is
that magnetic pitch angles are further affected by streaming motions
associated with the spiral pattern (Moss, 1998).

As shown by Moss et al.\ (2000), magnetic pitch angle can be affected
by an axisymmetric radial inflow (as well as outflow):
\[
\tan p
\simeq-\sqrt{\frac{R_\alpha}{|R_\omega|}} \left(1-\sfrac12 {\cal
L}\sqrt{\frac{\pi}{-D}}\,\right), \qquad {\cal L}=\frac{h^2}{2\beta}
\left(\frac{v_r}{r}-\frac{\partial v_r}{\partial r}\right),
\]
which is useful to compare with Eq.~(\ref{p}). This effect is
important if $v_r\ga2\beta/h\simeq1\kms$ (cf.\ Sect.~\ref{RMS}).

Twisting of a horizontal primordial magnetic field by galactic
differential rotation leads a tightly wound magnetic structure with
magnetic field direction alternating with radius at a progressively
smaller scale $\Delta r\simeq R/|G|t$ with $\tan p\simeq-(|G|t)^{-1}$,
where $R\simeq10\kpc$ is the scale of variation in $\Omega$ (see
Moffatt, 1978, Sect.\ 3.3; Howard \& Kulsrud, 1997; Kulsrud, 1999
for a detailed discussion).  The winding-up proceeds until a time
$t_0\simeq5\times10^9\yr$ such that $|G|t_0\simeq|C_\omega|^{1/2},$
where $C_\omega=GR^2/\beta=R_\omega R^2/h^2\simeq10^3$--$10^4$.  At
later times, the alternating magnetic field rapidly decays because of
diffusion and reconnection.  The resulting maximum magnetic field
strength achieved at $t_0$ is given by
\begin{equation}
B_{\rm max}\simeq B_0 |C_\omega|^{1/2}\;, \label{Bprim}
\end{equation}
where $B_0$ is the external magnetic field; the magnetic field
reverses at a small radial scale $\Delta r\simeq
R|C_\omega|^{-1/2}\simeq100\p$.  The magnetic pitch angle at $t_0$ is
of the order of $|p|\simeq |C_\omega|^{-1/2}\la1^\circ$, i.e., much
smaller than observed. This picture cannot be reconciled with
observations (cf.\ Kulsrud, 1999). It can be argued that streaming
motions could make magnetic lines more open and parallel to the
galactic spiral arms. However, then magnetic field will reverse on a
small scale not only along radius, but also along azimuth. Such
magnetic structures are quite different from what is observed.  The
moderate magnetic pitch angles observed in spiral galaxies are a
direct indication that the regular magnetic field is not frozen into
the interstellar gas and has to be maintained by the dynamo (Beck,
2000).

MHD density waves can produce non-zero magnetic pitch angle only in
non-axisymmetric magnetic structures (Lou \& Fan, 1998); an
axisymmetric MHD wave suggested to explain the magnetic ring in M\,31
(Lou \& Fan, 2000) can only have vanishing radial magnetic field,
$p\equiv0$.

%--------------------------------------------------------
\subsection{Even (quadrupole) symmetry of magnetic field
in the Milky Way}\label{QS}
One of the most convincing arguments in favour of the galactic dynamo
theory comes from the symmetry of the observed regular magnetic field
with respect to the Galactic equator in the Milky Way.  As shown by
wavelet analysis of the Faraday rotation measures of extragalactic
radio sources, the horizontal components of the local regular
magnetic field definitely have even parity being similarly directed
on both sides of the midplane (Frick et al., 2000b). This symmetry is
naturally explained by dynamo theory where even parity is strongly
favoured against odd parity because the even field has twice larger
scale in the vertical coordinate. Because of the difference in scale,
the even field is subject to weaker destruction by magnetic
diffusion, so the dynamo can regenerate even fields more efficiently
(see Ruzmaikin et al., 1988a,b).

Primordial magnetic field twisted by differential rotation can have
even vertical symmetry if it is parallel to the disk plane. However,
then the field is rapidly destroyed by twisting and reconnection as
described in Sect.~\ref{MPA}. If, otherwise, the primordial field is
parallel to the rotation axis and amplified by the vertical
rotational shear $\partial\Omega/\partial z$ (which is weak in
galaxies anyway), it can avoid catastrophic decay (Moffatt, 1978,
Sect.~3.11), but then it will have odd parity in $z$, which is ruled
out by the observed parity of the Milky Way field.

The derivation of the regular magnetic field of the Milky Way from
Faraday rotation measures of pulsars and extragalactic radio sources,
$\RM$, is complicated by the contribution of local magnetic
perturbations, so it is difficult to decide which features of the RM
sky are due to the regular magnetic field and which are produced by
localized magneto-ionic perturbations (e.g., supernova remnants).
Therefore, the same observational data have lead different authors to
different conclusions (see Frick et al., 2000b, for a recent review).
Odd parity of the Galactic magnetic field has been suggested by
Andreassian (1980, 1982) and, for the inner Galaxy, by Han et al.\
(1997). As stressed by Frick et al.\ (2000b), quantitative
methods of analysis are especially appropriate in this case.

Unfortunately, it is difficult to determine the parity of magnetic
fields in external galaxies. In galaxies seen edge-on, the disk is
depolarized, whereas Faraday rotation in the halo is weak. Beck et
al.\ (1994) found weak evidence of even magnetic parity in the
lower halo of NGC~253. The arrangement of polarization planes in the
halo of NGC~4631 (Beck, 2000) is very suggestive of odd
parity, but this does not exclude even parity in the disk.
In galaxies inclined to the line of sight, the amount of Faraday
rotation produced by an odd (antisymmetric) magnetic field differs
from zero because Faraday rotation and emission occur in the same
volume; as a result, emission originating at the far half of the
galactic layer will have small or zero net rotation, whereas emission
from the near half will have significant rotation produced by the
unidirectional magnetic field in that half.  Therefore, Faraday
rotation measures produced by even and odd magnetic structures of the
same strength only differ by a factor of two (Krause et al., 1989a;
Sokoloff et al., 1998) and it is difficult to distinguish between the
two possibilities.

An interesting method to determine the parity of magnetic field in an
external galaxy has been suggested by Han et al.\ (1998). These
authors note that the contribution of the galaxy to the $\RM$ of a
background radio source will be equal to the intrinsic $\RM$ of the
galaxy if the magnetic field has even parity. For odd parity, the
galaxy will not contribute to the $\RM$ of a background source,
whereas any intrinsic $\RM$ will remain. The implementation of the
method requires either a statistically significant sample of background
sources or an extended single background source.

%--------------------------------------------------------
\subsection{Azimuthal structure}
Non-axisymmetric magnetic fields in a differentially rotating object
are subject to twisting and enhanced dissipation as described in
Sect.~\ref{MPA}.  The dynamo can compensate for the losses, but
axisymmetric magnetic fields are still easier to maintain (R\"adler,
1986). A few lowest non-axisymmetric modes with azimuthal wave
numbers
\begin{equation}
m\la\frac{R}{h}|R_\omega|^{-1/4}\simeq2                \label{mmm}
\end{equation}
can be maintained in thin galactic disks where $h\ll R$ (Ruzmaikin et
al., 1988a, Sect.\ VI.7). However, the {\em dominance\/} of
non-axisymmetric modes in most galaxies would be difficult to
explain.

Early interpretations of Faraday rotation in spiral
galaxies indicated strong dominance of bisymmetric magnetic
structures $(m=1)$, $\vect{B}\propto\exp{i\phi}$ with $\phi$ the
azimuthal angle (Sofue et al., 1986), and this was considered as a
severe difficulty of the dynamo theory and an evidence of the
primordial origin of galactic magnetic fields. Despite effort, dynamo
models could not explain the apparent widespread dominance of
bisymmetric magnetic structures. However, what seemed to be a
difficulty of the dynamo theory has turned out to be its advantage as
observations with better sensitivity and resolution and better
interpretations have led to a dramatic revision of the observational
picture. The present-day understanding is that modestly distorted
axisymmetric magnetic structures occur in most galaxies, wherein the
dominant axisymmetric mode is mixed with weaker higher azimuthal
modes (Beck et al., 1996; Beck, 2000).  Among nearby galaxies, only
M\,81 remains candidate for a bisymmetric magnetic structure (Krause et
al., 1989b); the interesting case of M\,51 is discussed below. Deviations
from precise axial symmetry can result from the spiral pattern,
asymmetry of the parent galaxy, etc.  Dominant bisymmetric magnetic
fields can be maintained by the dynamo action near the corotation
radius due to a linear resonance with the spiral pattern (Mestel \&
Subramanian, 1991; Subramanian \& Mestel, 1993; Moss, 1996) or
nonlinear trapping of the field by the spiral pattern (Bykov et al.,
1997).

Twisting of a horizontal magnetic field by differential rotation
generally produces a bisymmetric magnetic field, $m=1$. A twisted
primordial magnetic field can result in an axisymmetric configuration
near the galactic centre if the initial state is asymmetric (Sofue
et al., 1986; Nordlund, 2000), with a maximum of the primordial field
displaced from the disk's rotation axis where the gas density is
normally maximum. The requirement that magnetic fields in most
spiral galaxies are axially symmetric within large radius (in fact,
the whole galaxy) would need a systematic strong asymmetry in the
initial state, with distinct initial distributions of magnetic field
and gas density, which would be difficult to explain.

%--------------------------
\subsubsection{M\,33}
A bisymmetric magnetic structure has been suggested for M\,33 by
Buczilowski \& Beck (1991). Recent results of Fletcher et al.\
(2000) indicate that magnetic field in M\,33 can represent an
axisymmetric structure distorted by a two-armed spiral pattern.  New
observations are needed to decide between these possibilities.  The
magnetic spiral, as well as the optical spiral pattern, is rather
open with $p\simeq-40^\circ$.

A dynamo model for M\,33 was suggested by Starchenko \& Shukurov
(1989) who obtained WKB asymptotics for the growth rates of
axisymmetric and non-axisymmetric modes of the mean-field
$\alpha\omega$-dynamo neglecting any nonlinear effects and the spiral
structure. They concluded that M\,33 is likely to support two lowest
magnetic modes ($m=1,\ 2$) because its
rotation is rather close to a rigid one.  More specifically, the
bisymmetric mode can grow in M\,33 provided $Q=|d\ln\Omega/d\ln
r|\,(v/l\Omega)(h/l)^2\la25$, which seems to be the case. As shown by
Eq.~(\ref{p}), weaker differential rotation with
$R_\alpha\simeq|R_\omega|$ naturally leads to a more open magnetic
spiral. These results need to be reconsidered and developed further
using more advanced dynamo models.

%-------------------------------------------------
\begin{figure}[tb]\label{M51}
 \begin{tabular}{cc}
\vspace*{-6mm}\psfig{figure=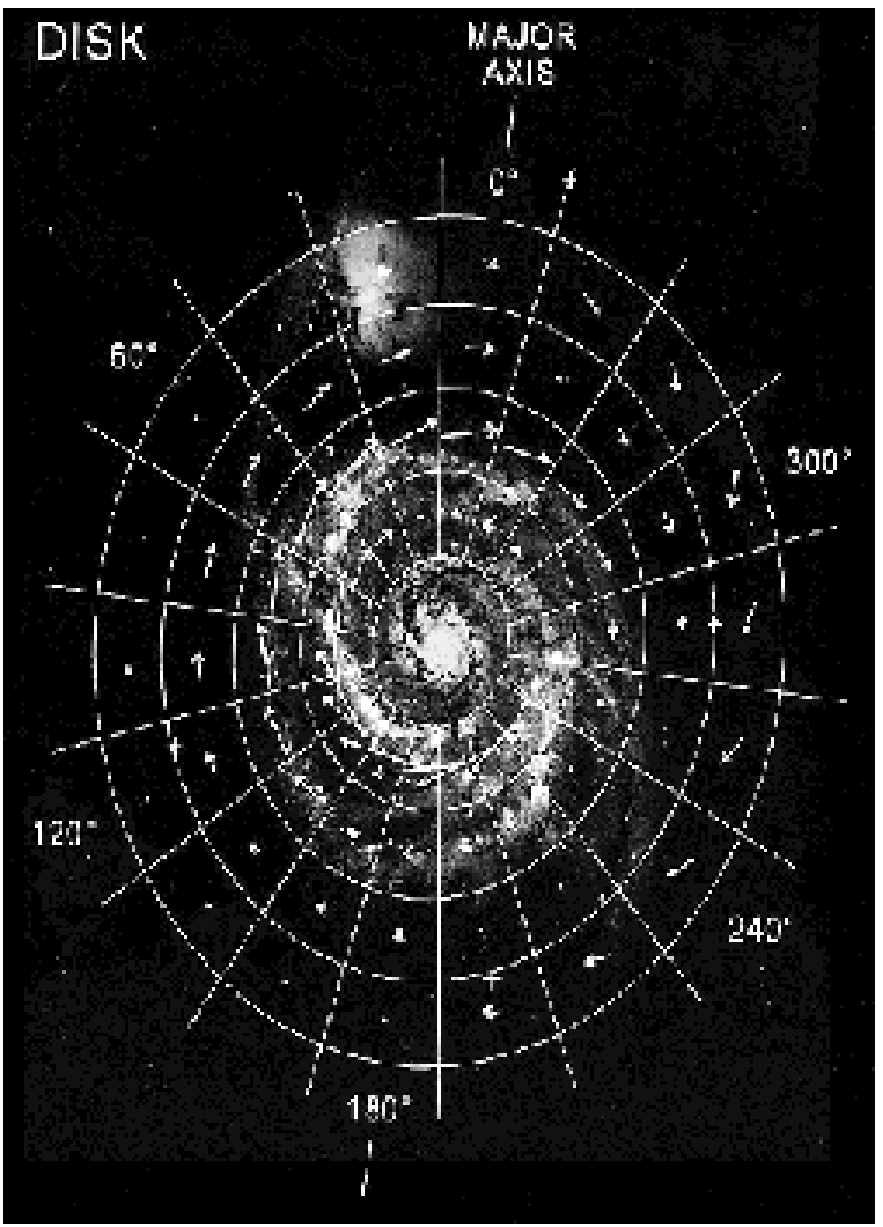,width=7.3truecm,height=8.62truecm,clip=}\vspace*{6mm}
     &
\psfig{figure=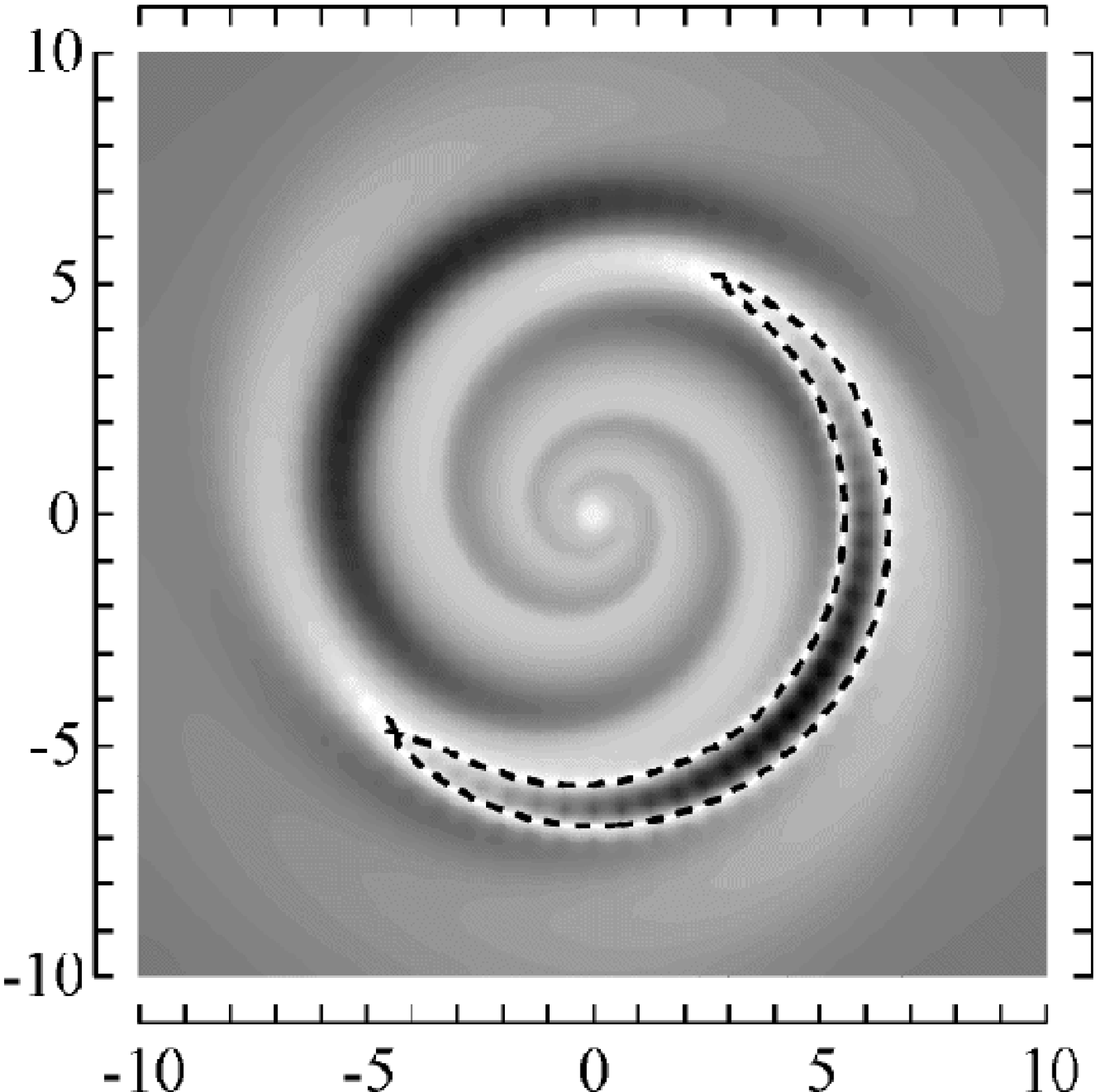,width=7.3truecm}
\end{tabular}
\caption[]{{\em Left panel:\/} Magnetic structure in the disk of M\,51
obtained from radio polarization observations at $\lambda\lambda2.8,
6.2, 18.0$ and $20.5\cm$ smoothed to a resolution of $\simeq3.5\kpc$;
 arrows indicate the direction and strength of the regular magnetic
field on the polar grid shown superimposed on the optical picture
(see Berkhuijsen et al., 1997, for detail).  The grid radii are 3, 6,
9, 12 and 15\,kpc. Note reversed magnetic field at azimuthal angles
$300^\circ$--0 in the inner ring.
{\em Right panel:\/}  Magnetic field strengths
from the nonlinear dynamo model of Bykov et al.\ (1997) is shown with
shades of grey (darker shade corresponds to stronger field). Magnetic
field is reversed within the zero-level contour shown dashed; scale
is given in kpc.  The model is based on the rotation curve of M\,51
(Tully, 1974), with the pitch angle of the spiral arms $-15^\circ$
and corotation radius 6\,kpc.  The magnetic structure rotates rigidly
together with the spiral pattern visible in the shades of grey.}
\end{figure}
%-----------------------------------------------------------------------------

%--------------------------------------------------------
\subsection{A composite magnetic structure in M\,51 and
magnetic reversals in the Milky Way}                        \label{cmsM51}
A striking example of a complicated magnetic structure that can
hardly be explained by any mechanism other than the dynamo is
provided by the galaxy M\,51 (Berkhuijsen et al., 1997).  As shown in
Fig.~2, the regular magnetic fields in the disk is reversed in a
region about 3 by 8\,kpc in size extended along azimuth at
galactocentric radii $r=3$--6\,kpc.  A significant deviation from
axial symmetry in the disk has been detected out to $r=9\kpc$,
although it is too weak to result in magnetic field reversal.
Furthermore, the regular magnetic field in the halo of M\,51 has a
structure distinct from that in the disk --- the halo field is nearly
axisymmetric and directed oppositely to that in the disk in most of
the galaxy. An external magnetic field should have a rather peculiar
form to be twisted into such a configuration!

A nonlinear dynamo model based on the rotation curve of M\,51,
developed by Bykov et al.\ (1997), shows that a region with reversed
magnetic field can occur in the disk near the corotation
radius of the spiral pattern. Near the corotation, non-axisymmetric
(bisymmetric) magnetic field can be trapped by the spiral pattern and
maintained over the galactic lifetime. The effect is favoured by a
smaller pitch angle of the spiral arms, thinner gaseous disk, weaker
rotational shear and stronger spiral pattern. A disk dynamo solution
of Bykov et al.\ (1997) arguably similar to the structure observed in
M\,51 is shown in Fig.~2.

Distinct azimuthal magnetic structures in the disk and the halo can
be readily explained by dynamo theory as non-axisymmetric magnetic
fields can be maintained only in the thin disk but not in the
quasispherical halo where $h\simeq R$ and $|R_\omega|\gg1$ in
Eq.~(\ref{mmm}).  Moreover, dynamo action in the disk and the halo can
proceed almost independently of each other producing distinctly
directed magnetic fields (Sokoloff \& Shukurov, 1990).

%--------------------------------------------------------
%\subsubsection{Milky Way}\label{MW}
Another case of a regular magnetic field with unusual radial
structure is the Milky Way where magnetic field reversals
are observed in the inner Galaxy between the Orion and Sagittarius
arms at $r\approx7.9\kpc$  (Simard-Normandin \& Kronberg, 1980; Rand
\& Kulkarni, 1989; Rand \& Lyne, 1994; Frick et al., 2000b) and,
possibly, in the outer Galaxy between the Orion and Perseus arms at
$r\approx10.5\kpc$  (Agafonov et al., 1988; Frick et al., 2000b; see,
however, Vall\'ee, 1983). The reversals were first interpreted as an
indication of a global bisymmetric magnetic structure (Sofue \&
Fujimoto, 1983), but it has been shown that dynamo-generated
axisymmetric magnetic field can have reversals at the appropriate scale
(Ruzmaikin et al., 1985; Poezd et al., 1993). Both interpretations
presume that the reversals are of a global nature, i.e., they extend
over the whole Galaxy to all azimuthal angles (or radii in the case of
the bisymmetric structure). This leads to a question why reversals at
this radial scale are not observed in any other galaxy (Beck, 2000).
Poezd et al.\ (1993) argue that the lifetime of the reversals is
sensitive to subtle features of the rotation curve and the geometry of
the ionized gas layer (see also Belyanin et al., 1993) and demonstrate
that they are more probable to survive in the Milky Way than in M\,31.

However, the observational evidence of the reversals is restricted to
a relatively small neighbourhood of the Sun, of at most 3--5\,kpc
along azimuth. It is therefore quite possible that the reversals
are local and arise from a magnetic structure similar to that in the
disk of M\,51 as shown in Fig.~2. The reversed field in the Solar
neighbourhood has the same radial extent of 2--3\,kpc as in M\,51 and
occurs near the corotation radius. This possibility has not yet been
explored; its observational verification would require careful
analysis of pulsar Faraday rotation measures.

%--------------------------------------------------------
\subsection{The radial magnetic structure in M\,31}\label{RMS}
%--------------------------------------------------------
%\subsubsection{The Andromeda nebula}
An important evidence in favour of galactic dynamos is the magnetic ring
in M\,31 (Beck, 1982), predicted by dynamo theory (Ruzmaikin \&
Shukurov, 1981). The dynamo model of Ruzmaikin \& Shukurov (1981) was
based on the double-peaked rotation curve of Rubin \& Ford (1970),
Deharveng \& Pellet (1975) and Haud (1981) shown in Fig.~1, where
rotational shear is strongly reduced at $r=2$--6\,kpc. As a result,
$R_\omega$ is small and even positive in this radial range, so $|D|\la
D_{\rm cr}$ and the dynamo cannot maintain any regular magnetic
field at  $r=2$--6\,kpc.

An attractive aspect of this theory is that both magnetic and gas
rings are attributed to the same feature of the rotation curve.
Angular momentum transport by viscous stress leads to matter inflow
at a rate $\dot
M=2\pi\sigma\nu(r/\Omega)\,d\Omega/dr\simeq0.1\,M_\odot\yr^{-1}$,
where $\nu\simeq\beta$ is the turbulent viscosity, resulting in the
radial inflow at a speed $v_r=\dot M/2\pi r\sigma$ with $\sigma$ the
gas surface density.  In the nearly-rigidly rotating parts, $v_r$ is
reduced and matter piles up outside this region producing gas ring.
Gravitational torques from spiral arms can further enhance the inflow
(see Moss et al., 2000, for a discussion), so the total radial
velocity is expected to be of order $1\kms$ at $r=10\kpc$.

The double-peaked rotation curve of M\,31 is consistent with the
existence of  both magnetic and gas rings. The situation is different
with the more recent rotation curve of Braun (1991) which does
not have a double-peaked shape (Fig.~1). The difference between the
two rotation curves arises mainly from the fact that Braun allows for
significant displacements of spiral arm segments from the galactic
midplane: this results in a revision of the segments' galactocentric
distances for regions away from the major axis. We note that the CO
velocity field at the major axis (Loinard et al., 1995) is compatible
with a double-peaked rotation curve.

%-------------------------------------------------
\begin{figure}[tb]\label{M31d}
 \begin{tabular}{cc}
\psfig{figure=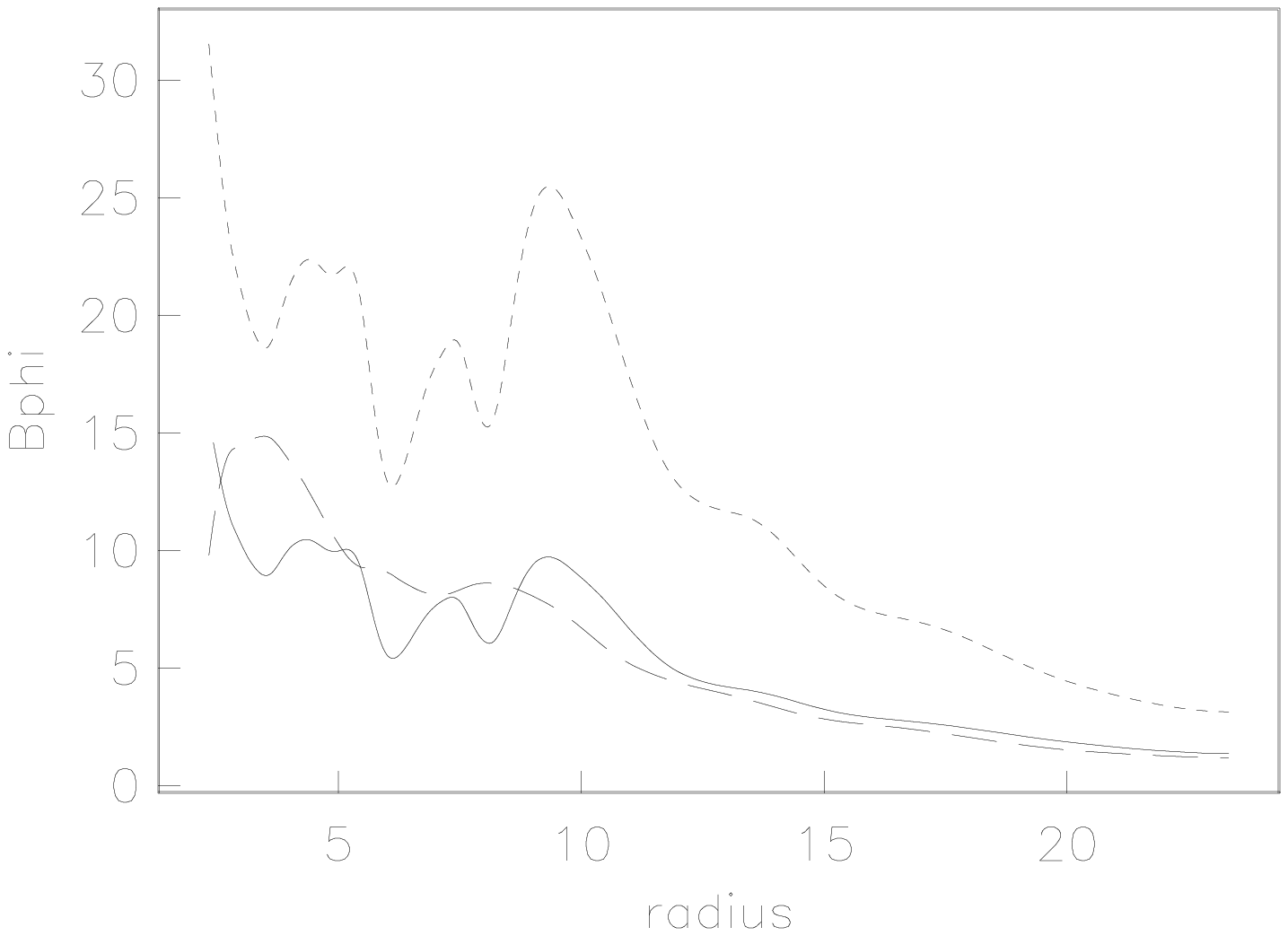,width=6.9truecm}
     &
\psfig{figure=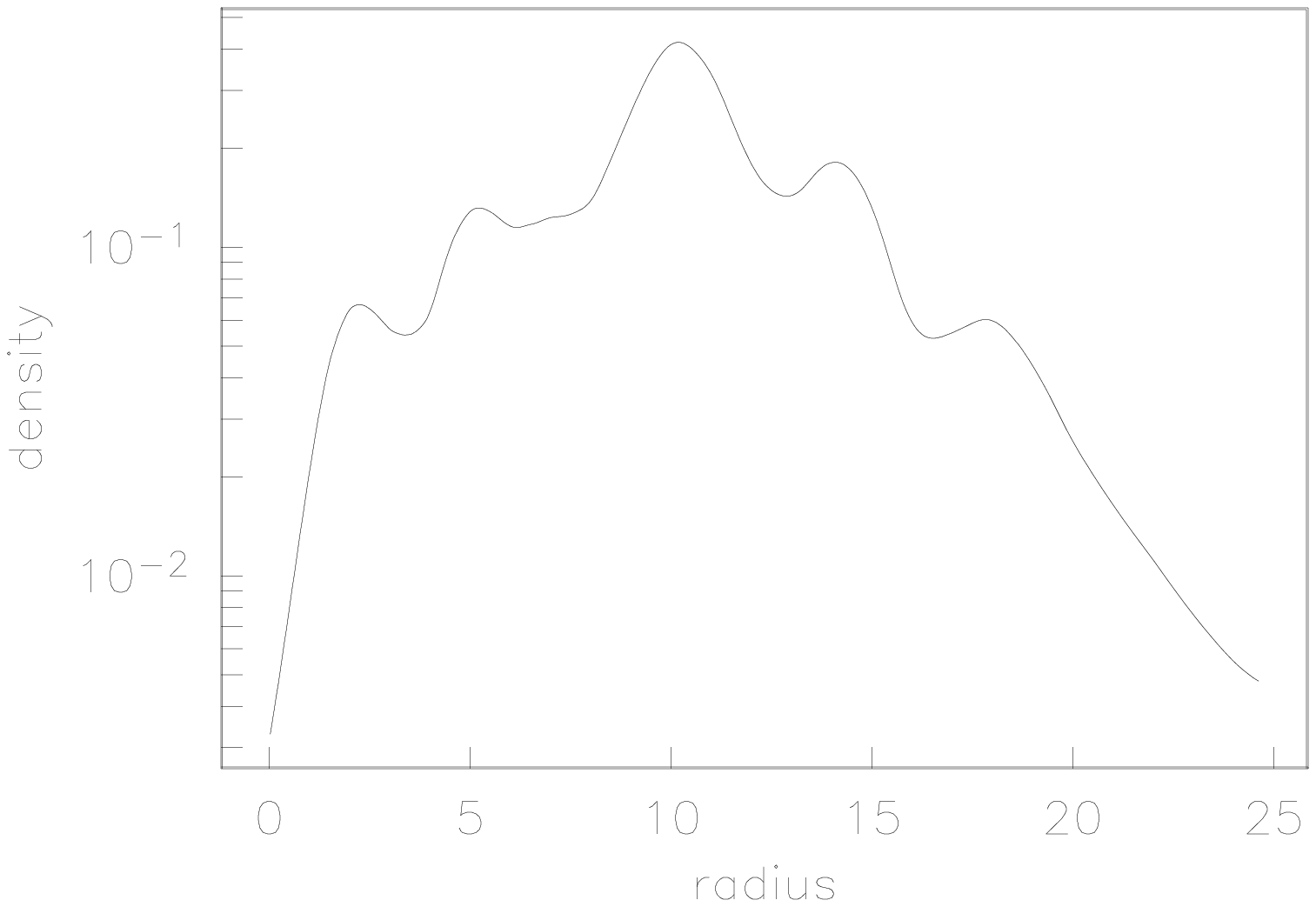,width=7.3truecm}
\end{tabular}
\caption[]{{\em Left panel:\/} The radial profiles of the regular
azimuthal magnetic field in M\,31 obtained from various versions of the
dynamo model of Moss et al.\ (1998) based on the rotation curve of
Braun (1991). {\em Right panel:\/} The radial profile of gas number
density adopted by Moss et al. In both panels, radius is given in
kpc, magnetic field in the units of $0.26\mkG$, and density in
cm$^{-3}$. The magnetic ring is only pronounced in the model shown with
short-dashed line which corresponds to relatively strong dynamo
action ($R_\alpha$ enhanced roughly by a factor of 6 above the
critical value). The radial variations in the magnetic field
strength are largely controlled by those in gas density.  Results
inside the 2\,kpc radius, where the Braun's results are unreliable,
have been distorted by modifications of the rotation curve and
density profile needed for technical reasons.}
\end{figure}
%-----------------------------------------------------------------------------

With Braun's rotation curve, the magnetic field can
concentrate into a ring mainly because the gas is in the ring and
$B\propto\rho^{1/2}$ as shown in Eq.~(\ref{B}).  The dynamo model of
Moss et al.\ (1998) based on the rotation curve of Braun (1991) has
difficulties in reproducing a magnetic ring as well pronounced as
implied by the observed amount of Faraday rotation --- see Fig.~3.
This has lead to an idea that magnetic field can be significant at
$r=2$--$6\kpc$ in M\,31. This has prompted Han et al.\ (1998) to search
for magnetic fields at $r=2$--6\,kpc that could have escaped detection
because of reduced density of cosmic ray electrons at those radii.
These authors have found that two out of three background polarized
radio sources seen through that region of M\,31 have Faraday rotation
measures compatible with the results of Moss et al.\ (1998) shown
with solid and long-dashed lines in Fig.~3.  They further conclude
that this indicates an even symmetry of the regular magnetic field.
This is encouraging, but a statistically representative sample of
background sources has to be used to reach definite conclusions
because of their unknown intrinsic $\RM$.

%Circumstantial evidence of genuine magnetic ring in M\,31 is provided
%by magnetic pitch angles at $r\la10\kpc$ shown in Fig.~1 where the
%observed values seem to be closer to the estimate of Eq.~(\ref{p})
%arising for a weak magnetic field than that of Eq.~(\ref{pn}) valid
%for a strong field. However, definite conclusion should await more
%refined dynamo models that include radial inflow.

With a double-peaked rotation curve, a primordial magnetic field with a
uniform radial component could have been twisted to produce a
magnetic ring by virtue of Eq.~(\ref{Bprim}). In this case the
primordial and dynamo theories have similar problems and
possibilities regarding the magnetic ring in M\,31.

Lou \& Fan (2000) attribute the magnetic ring in M\,31 to an
axisymmetric mode of MHD density waves. Because of the axial symmetry
of the wave, the magnetic field in the ring must be purely azimuthal,
$p=0$, in contrast to the observed structure with a significant
pitch angle (Fig.~1). Furthermore, the ring can hardly
represent a wave packet as envisaged in this theory because then its
group velocity must be comparable to the Alfv\'en speed of $30\kms$ (Lou
\& Fan, 1998) and so the ring should be travelling at this speed
along radius to traverse 30\,kpc in $10^9\yr$. The implication would
be that the ring is a transient with a short lifetime of order
$3\times10^8\yr$. And, of course, the theory cannot explain the
origin of an azimuthal magnetic field required to launch the wave
packet.

%--------------------------------------------------------
\subsection{Strength of the regular magnetic field}\label{SRMF}
Interstellar regular magnetic fields are close to energy
equipartition with interstellar turbulence. This directly indicates
that the regular magnetic field is coupled to the turbulent gas
motions. (Note that $l\Omega$ does not differ much from the turbulent
velocity $v$ in Eq.~(\ref{B}).)  To appreciate the importance of this
conclusion, consider primordial magnetic field twisted by
differential rotation. Its maximum strength given by
Eq.~(\ref{Bprim}) as $B_{\rm max}\simeq 10^2B_0$ is controlled by the
strength of the primordial field $B_0$, and so this theory, if
applicable, would result in stringent constraints on extragalactic
magnetic fields.

The theory of MHD density waves relates magnetic field excess in spiral
arms to the enhancement in stellar density, $\Delta B_{\rm
arm}/\langle B\rangle=\Delta \Sigma_{\rm arm}/\langle\Sigma\rangle$
(Lou \& Fan, 1998), where $\Sigma$ is the stellar surface density
and angular brackets denote azimuthal averaging. Arm intensities in
magnetic field and stellar surface density in NGC~6946 have been
estimated by Frick et al.\ (2000a) who applied wavelet transform
techniques to radio polarization maps at $\lambda\lambda3.5$ and
$6.2\cm$ and to the galaxy image in broadband red light.  Their
results indicate that the mean relative intensity of magnetic spiral
arms remains rather constant with galactocentric radius at a level of
0.3--0.6. On the contrary, the relative strength of the stellar arms
systematically grows with radius from very small values in the inner
galaxy to 0.3--0.7 at $r=5$--6\,kpc, and then decreases to remain at
a level of 0.1--0.3 out to $r=12\kpc$. The distinct magnitudes and
radial trends in the strengths of magnetic and stellar arms in
NGC~6946 do not seem to support the idea that the magnetic arms are due to MHD density
waves.

%---------------------------------------------------
%\section{Magnetic field in M\,33}\label{M33}

\section*{Acknowledgments}
I am grateful to R.~Beck, E.M.~Berkhuijsen, A.~Brandenburg, J.~Lequeux,
D.D.~Sokoloff, K.~Subramanian and R.A.M.\ Walterbos for helpful
comments and discussions. This work has been supported by  NATO
Collaborative Linkage Grant PST.CLG~974737. I am grateful to the {\em
Wilhelm und Else Heraeus-Stiftung\/} for financial support.

%------------------------------------------------------
\section*{References}\noindent

\references

Andreassian R.R. \Journal{1980}{{\em Astrofizika}}{16}{707}.

Andreassian R.R. \Journal{1982}{{\em Astrofizika}}{18}{255}.

Agafonov G.I., Ruzmaikin A.A., Sokoloff D.D.
\Journal{1988}{\SA}{32}{268}.

Beck R. \Journal{1982}{\AAp}{106}{121}.

Beck R. \Journal{2000}{{\em Phil.\ Trans.\ R.\ Soc.\ Lond}}{358}{777}.

Beck R., Carilli C.L., Holdaway M.A., Klein U.
\Journal{1994}{\AAp}{292}{409}.

Beck R., Brandenburg A., Moss D., Shukurov A., Sokoloff D.
\Journal{1996}{\ARAaAp}{34}{155}.

Berkhuijsen E.M., Horellou C., Krause M., Neininger N., Poezd A.D.,
Shukurov A., Sokoloff D.D. \Journal{1997}{\AAp}{318}{700}.

Beck R., Ehle M., Shoutenkov V., Shukurov A., Sokoloff D.
\Journal{1999}{\Nat}{397}{324}.

Belyanin M., Sokoloff D., Shukurov A. \Journal{1993}{{\em Geophys.\
Astrophys.\ Fluid Dyn.}}{68}{237}.

Brandenburg A. (2000) {\ApJ,} submitted (astro-ph/0006186).

Braun R. \Journal{1991}{\ApJ}{372}{54}.

Buczilowski U.R., Beck R. \Journal{1991}{\AAp}{241}{47}.

Bykov A., Popov V., Shukurov A., Sokoloff D.
\Journal{1997}{\MNRAS}{292}{1}.

Deharveng J.M., Pellet A. \Journal{1975}{\AAp}{38}{15}.

Fletcher A., Beck R., Berkhuijsen E.M., Shukurov A. (2000), this
volume.

Frick P., Beck R., Shukurov A., Sokoloff D., Ehle M., Kamphuis
J. (2000a) {\MNRAS}, in press.

Frick P., Stepanov R., Shukurov A., Sokoloff D. (2000b) {\em Mon.\
Not.\ R.~Astron.\ Soc.}, submitted.

Han J.L., Manchester R.N., Berkhuijsen E.M., Beck R.
\Journal{1997}{\AAp}{322}{98}.

Han J.L., Beck R., Berkhuijsen E.M. \Journal{1998}{\AAp}{335}{1117}.

Haud U. \Journal{1981}{\em Astrophys.\ Space Sci.}{76}{477}.

Howard A.M., Kulsrud R.M. \Journal{1997}{\ApJ}{483}{648}.

Krause M., Hummel E., Beck R. \Journal{1989a}{\AAp}{217}{4}.

Krause M., Beck R., Hummel E. \Journal{1989b}{\AAp}{217}{17}.

Krasheninnikova Y., Ruzmaikin A., Sokoloff D., Shukurov A.
\Journal{1989}{\AAp}{213}{19}.

Kulsrud R.M. \Journal{1999}{\ARAaAp}{37}{37}.

Loinard L., Allen R.J., Lequeux J. \Journal{1995}{\AAp}{301}{68}.

Lou Y.-Q., Fan A. \Journal{1996}{\Nat}{383}{800}.

Lou Y.-Q., Fan A. \Journal{1998}{\MNRAS}{297}{84}.

Lou Y.-Q., Fan A. (2000) {\em Mon.\ Not.\ R.~Astron.\ Soc.}, in press.

Mestel L., Subramanian K. \Journal{1991}{\MNRAS}{248}{677}.

Moffatt H.K. (1978) {\em Magnetic Field Generation in Electrically
Conducting Fluids,} Cambridge Univ.\ Press.

Moss D. \Journal{1996}{\AAp}{308}{381}.

Moss D. \Journal{1998}{\MNRAS}{297}{860}.

Moss D., Shukurov A., Sokoloff D., Berkhuijsen E.M., Beck R.
\Journal{1998}{\AAp}{335}{500}.

Moss D., Shukurov A., Sokoloff D. \Journal{1999}{\AAp}{343}{120}.

Moss D., Shukurov A., Sokoloff D. \Journal{2000}{\AAp}{358}{1142}.

Nordlund \AA. (2000) in Proc.\ JD14, IAU 24th Gen.\ Assembly, {\em Vistas Astron.}, in
press.

Parker E.N. \Journal{1992}{\ApJ}{401}{137}.

Poezd A., Shukurov A., Sokoloff D. \Journal{1993}{\MNRAS}{264}{285}.

R\"adler K.-H. (1986) in {\em Plasma Astrophysics}, ESA SP-251,
p.~285.

Rand R.J., Kulkarni S.R. \Journal{1989}{\ApJ}{343}{760}.

Rand R.J., Lyne A.G. \Journal{1994}{\MNRAS}{268}{497}.

Rubin V.C., Ford W.K. \Journal{1970}{\ApJ}{159}{379}.

Ruzmaikin A.A., Shukurov A.M. \Journal{1981}{\SA}{25}{553}.

Ruzmaikin A.A., Sokoloff D.D., Shukurov A.M.
\Journal{1985}{\AAp}{148}{335}.

Ruzmaikin A.A., Shukurov A.M., Sokoloff D.D. (1988a) {\em Magnetic
Fields of Galaxies,} Kluwer, Dordrecht.

Ruzmaikin A., Sokoloff D., Shukurov A. \Journal{1988b}{\Nat}{336}{341}.

Shukurov A. \Journal{1998}{\MNRAS}{299}{L21}

Shukurov A., Sokoloff D. \Journal{1998}{{\em Studia Geophys.\
Geod.}}{42}{391}.

Simard-Normandin M., Kronberg P.P. \Journal{1980}{\ApJ}{242}{74}.

Sofue Y., Fujimoto M. \Journal{1983}{\ApJ}{265}{722}.

Sofue Y., Fujimoto M., Wielebinski R.
\Journal{1986}{\ARAaAp}{24}{459}.

Sokoloff D., Shukurov A. \Journal{1990}{\Nat}{347}{51}.

Sokoloff D.D., Ruzmaikin A.A., Shukurov A. (1990) in {\em Galactic and
         Intergalactic Magnetic Fields.  IAU Symp.\ 140}, eds.\
         R.~Beck, P.P.~Kronberg and R.~Wielebinski, Kluwer,
         Dordrecht, p.\ 499.

Sokoloff D.D., Bykov A.A., Shukurov A., Berkhuijsen E.M., Beck R.,
Poezd A.D. \Journal{1998}{\MNRAS}{299}{189} (Erratum {\bf 303}, 207).

Starchenko S.V., Shukurov A.M. \Journal{1989}{\AAp}{214}{47}.

Subramanian K. \Journal{1999}{{\em Phys.\ Rev.\ Lett.}}{83}{2957}.

Subramanian K., Mestel L.  \Journal{1993}{\MNRAS}{265}{649}.

Tully R.B. \Journal{1974}{\ApJSS}{27}{437}.

Vall\'ee J.P. \Journal{1983}{\AAp}{124}{1983}.

Wielebinski R., Krause F. \Journal{1993}{\AAp\ Rev.}{4}{449}.

\end{document}